\documentclass[aps, pra, twocolumn, notitlepage, nofootinbib, longbibliography]{revtex4-1}

\usepackage{graphicx,amsmath,pstricks,float,subcaption,mathtools}

\begin{document}

\title{A causal limit to communication within an expanding cosmological civilization}

\author{S. Jay Olson}
 \email{stephanolson@boisestate.edu}
 \affiliation{Department of Physics, Boise State University, Boise, Idaho 83725, USA}
 
\date{\today}

\begin{abstract}
If a civilization embarks on high-speed intergalactic expansion, growing to a cosmological scale over time, communication between remote galaxies in the civilization will incur an extreme time delay, due to the distance involved. Indeed, if the net expansion speed $v$ is more than $\approx .26 \, c$, most of the final volume of such a civilization will not be able to signal the home galaxy at all, due to the presence of a causal horizon. We illustrate the regions of such a civilization according to the degree of ``conversation'' that is possible with the home galaxy, and describe how the geometry depends on expansion speed. We conclude by reflecting on the value of space settlement beyond the horizon, where colonies can never be observed by the initiating home galaxy.
\end{abstract}

\maketitle

\section{Introduction}

If a few enabling technologies are feasible, and an advanced civilization seeks to maximize its access to physical resources, the result is an expanding cosmological civilization (ECC) --- a wave of colonization that begins at a home galaxy and expands at high speed in all directions, saturating every galaxy along the way. Due to the large scale homogeneity of the universe, the geometry of such a civilization is an expanding sphere.

These theoretical civilizations have mostly been studied at a universe-wide scale, in order to quantify how they could come to fill the cosmos~\cite{olson2014}, their observable geometry from the vantage point of the Earth~\cite{olson2016,ord2021edges}, and to enable anthropic reasoning~\cite{olson2017a,olson2020,Hanson_2021,olson2021,cook2022}.

Here, we quantify a general \emph{internal} property of an ECC --- a fundamental limit to communication, imposed by causality. Since the spatial size of an ECC grows to become cosmological, so too will the time lag for communication. And because the growth of an ECC is ultimately limited by a cosmological horizon, communication will be as well.

The quantity of interest to us here is the \emph{degree of conversation}, $n$, that is possible between a colony and the home galaxy of the ECC. That is, the number of sequential, back-and-forth signals that are possible before the end of time. This will divide the ECC into spatial regions, corresponding to different values of $n$ (see figures 1 and 2).

The relative size of each region will depend on $v$, the speed of growth of the ECC. For a very high expansion speed, close to $v=c$, the large majority of the final ECC volume corresponds to the $n=0$ region, where colonies can send no signal at all back to the home galaxy. For lower expansion speeds, the relative size of the $n=0$ region is smaller.

We organize the paper as follows: Section II reviews the assumptions going into an ECC, while our main results are derived and illustrated in section III. The prospect of initiating a civilization, the bulk of which we can never hope to see, raises value questions that we conclude with in section IV.

\section{Background Assumptions and the Feasibility of an ECC}

Our results concern limits to communication within an expanding cosmological civilization, throughout the indefinite future.

This can be studied in a variety of contexts, but we will use a set of ``default assumptions'' that seem most plausible from our present-day point of view of physics and cosmology.

\begin{enumerate}
	\item We assume that standard-model physics is at least approximately true. In particular, we assume that the speed of a communication signal is equal to the speed of light, and that space flight is limited to some $v < 1$ (we use units in which $c = 1$). 
	\item We assume the standard cosmology ($\Lambda$CDM) will remain at least approximately valid over the epoch in which life is active. The most significant feature of this cosmology, for our purposes, is the existence of the cosmological horizon. The full Friedmann-Robertson-Walker solution only enters our result through a single number, namely the coordinate distance to the cosmological horizon at the present time. We denote this distance as $R_{max} \approx 16.6 \, Gly$ and the scale factor as $a(t)$. 
	\item We assume that the spherical front of intergalactic expansion proceeds at some constant speed $v$ in the comoving frame. This simplifying assumption can be justified with a simulated distribution of galaxies, and a plausible model for expansion~\cite{olson2018}. For example, assume each newly-colonized galaxy launches a new spacecraft to a speed $v$ on a geodesic flight toward every galaxy within a comoving range $r$. Then, for a wide range of $\{ v, r\}$, the net expansion speed of the spherical front formed by these galaxy-to-galaxy hops is approximately ``constant-$v$ in the comoving frame.''
	\item We focus on a single expanding civilization whose growth is not limited by collisions~\cite{olson2018,Hanson_2021} with other expanding civilizations. We do this to present a clean, limiting case that does not depend on alien life, though such collisions are possible and even quite likely, depending on how persuasive one finds the current state of anthropic reasoning~\cite{olson2020,Hanson_2021,cook2022}.
\end{enumerate} 

Such an extreme cosmological project does not require esoteric technologies relying on speculative physics. What is required is self-replication~\cite{freitas1980} and high-speed, long-duration space flight~\cite{fogg1988,sandberg2018}. Self-replicating intelligence is a practical process, familiar from biology, while spacecraft with a speed of $.2 c$ on a realistic budget are already in development~\cite{parkin2018}. 

An ECC does not require sending human bodies through intergalactic space, which would be enormously expensive and impractical. By contrast, sending the information required to reconstruct human bodies or minds at the destination should not be a limiting factor~\footnote{The information content of a human brain is often estimated to be on the order of hundreds of terabytes (give or take an order of magnitude), but the information density of present-day DNA storage techniques is hundreds of petabytes per gram~\cite{erlich2017}.}.

\section{Communication within an Expanding Cosmological Civilization}

Moving through space from cosmic time $t_i$ to time $t_f$ with constant speed $v$ (in the comoving frame of reference), the comoving coordinate distance traveled is:

\begin{eqnarray}
R = \int_{t_{i}}^{t_{f}} \, \frac{v}{a(t)} \, dt.
\end{eqnarray} 

Thus, starting from today (cosmic time $t_0$), the maximum coordinate distance possible to travel, moving at $v=1$ (the speed of light) is\footnote{We have used the solution fixed by $\Omega_{\Lambda 0}=.692$, $\Omega_{r0}=9 \times 10^{-5}$, $\Omega_{m0}=1-\Omega_{r0} -\Omega_{\Lambda 0}$, $H_0 =.069 \, Gyr^{-1}$.}:

\begin{eqnarray}
R_{max} = \int_{t_{0}}^{\infty} \, \frac{1}{a(t)} \, dt \approx 16.6 \, Gly. 
\end{eqnarray} 

The spatial region bounded by $R_{max}$ around us has been called the \emph{affectable universe} by Ord~\cite{ord2021edges}. Here, we will think of $R_{max}$ as a kind of ``travel distance budget,'' that will be consumed by sequential stages of a mission. 

$R_{max}$ is consumed by processes that take time. Either the time in which a spacecraft is traveling, or by back-and-forth communication (traveling light signals) between a colony and the home galaxy. Generally, the distance consumed from $R_{max}$ by any process is the distance light \emph{could have traveled} during that process.  

A simplified mission profile looks like this: At $t_0$, a spacecraft is launched, traveling at some average speed $v < 1$. It eventually reaches its destination galaxy at comoving coordinate distance $R$, and establishes a colony there. The colony then signals the home galaxy, which sends a reply, with the colony replying to the reply, and so on. This communication results in $n$ sequential trips of comoving coordinate distance $R$, at the speed of light ($v=1$).

Define $R_n$ to be the coordinate distance where exactly $n$ sequential communication signals are possible between the newly-established colony and the home galaxy, with no time left over. This signaling will consume $n R_n$ from the distance budget. The initial stage of spacecraft travel to establish the colony also consumes $R_{n}/v$ from the budget, i.e. the distance light \emph{could have traveled} during the time the spacecraft was traveling at speed $v$. Thus, by definition of $R_n$, the distance budget equation is:

\begin{eqnarray}
\frac{R_{n}}{v} + n \, R_{n} = R_{max}.
\end{eqnarray}       

Or, solving for $R_n$:

\begin{eqnarray}
R_n = \frac{v \, R_{max}}{1 + v \, n}.
\end{eqnarray}

The distance $R_0 = v \, R_{max}$ is the coordinate distance eventually reached by a spacecraft that simply travels forever at speed $v$. The distance $R_1$ is the greatest distance at which a colony can still signal the home galaxy, after being established. $R_2$ is the greatest distance at which a colony can both signal the home galaxy and receive a reply, etc. etc. etc.

We thus call $n$ the \emph{degree of conversation} that is possible between any colony and the home galaxy.

If we imagine not just a single spacecraft, but a spherical front of colonization that expands from the home galaxy at average speed $v$, the resulting ECC is divided into regions by spheres of radius $R_n$ (see figures 1 and 2). The fraction of comoving volume (and thus matter or colonies) of the final civilization that exists within $R_n$ is thus:

\begin{eqnarray}
\frac{V_n}{V_0} =  \frac{1}{(1 + v \, n)^3}.
\end{eqnarray} 

At a net expansion speed of $v = 2^{1/3} - 1 \approx .26$, half of the final volume of such a civilization will lie beyond $R_1$, and thus not be able to signal the home galaxy (i.e. the home galaxy can never see the $n=0$ region become established). In the limit of a colonizing spacecraft front that approaches $v = 1$, the $n=0$ region would correspond to 7/8 of the final volume of the civilization. 

This result is a fundamental barrier of causality --- colonies beyond $R_1$ are established at a time when the coordinate distance to the cosmological event horizon has become shorter than $R_1$, due to the cosmic acceleration. From the point of view of the home galaxy, colonizing the region beyond $R_1$ is much like colonizing the interior of a black hole.

The degree of conversation within gravitationally bound systems (e.g. individual galaxy groups and clusters) is not limited by the expansion of the universe, in the standard $\Lambda CDM$ cosmology. But all systems that are not gravitationally bound to our Local Group will eventually be pushed away, resulting in a bounded $n$, relative to the Milky Way. For example, the nearest large galaxy outside our Local Group is NGC 300, about 6 Mly away~\cite{rizzi2006}. We can solve equation 3 for $n$, giving:

\begin{eqnarray}
n= \frac{R_{max}}{R_n} - \frac{1}{v}.
\end{eqnarray}  

If we take $R_{max} = 16.6$ Gly, and $R_n = 6$ Mly, the degree of conversation $n$ is approximately $R_{max}/R_{n} = 2770$, for any plausible value of $v$. Since this is the nearest unbound galaxy, $\approx 2770$ represents an approximate upper bound on the degree of conversation between the Milky Way and any possible settlement beyond our Local Group. 

This number comes down quickly as one examines the local structure of the universe. The nearest distinct galaxy group, the IC 342 group, at a distance of roughly 11 Mly~\cite{anand2019}, yields a degree of conversation of approximately 1510 for plausible $v$. The nearest major supercluster (beyond Laniakea) is the Perseus–Pisces supercluster, which at 230 Mly yields a degree of conversation around 70.

\begin{widetext}
	
	\begin{figure}
		\centering
		\includegraphics[width=1.0\textwidth]{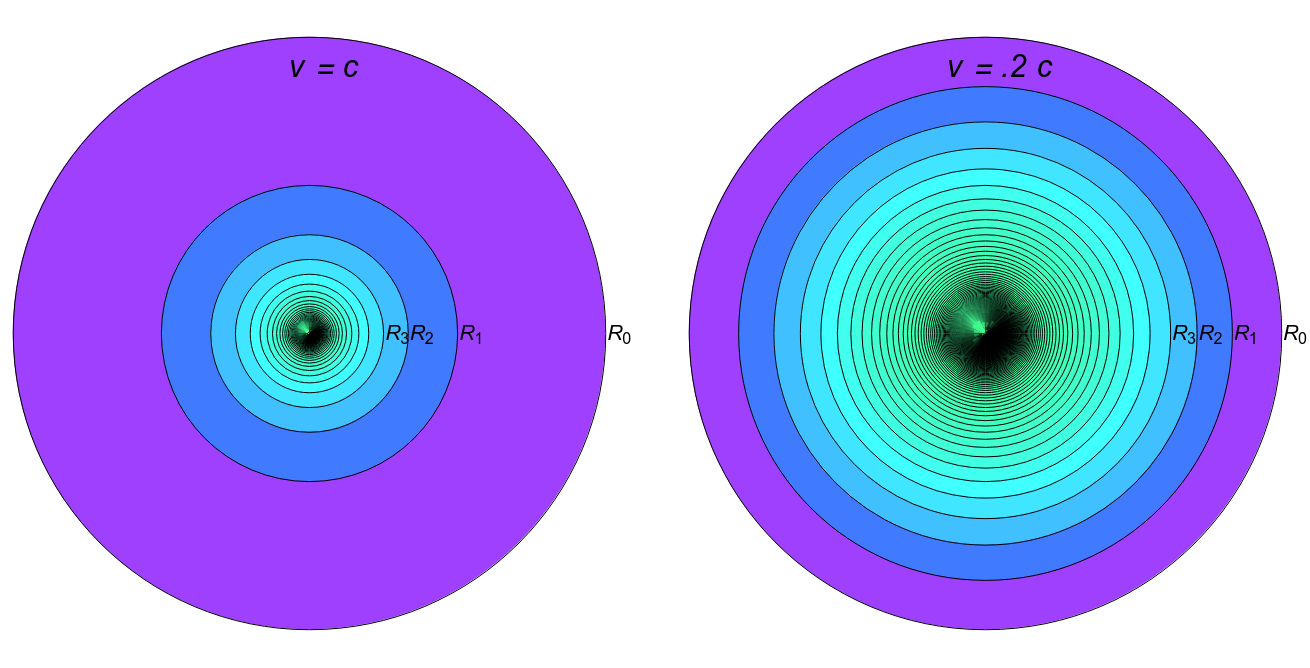}
		\caption{The relative size of communication regions for $v=1$ and $v=0.2$, with the final domain scaled to equal size, for comparison. High expansion velocity means that far more of the final domain exists beyond $R_1$, and will never be able to signal the home galaxy. }
	\end{figure}

\begin{figure}
	\centering
	\includegraphics[width=1.0\textwidth]{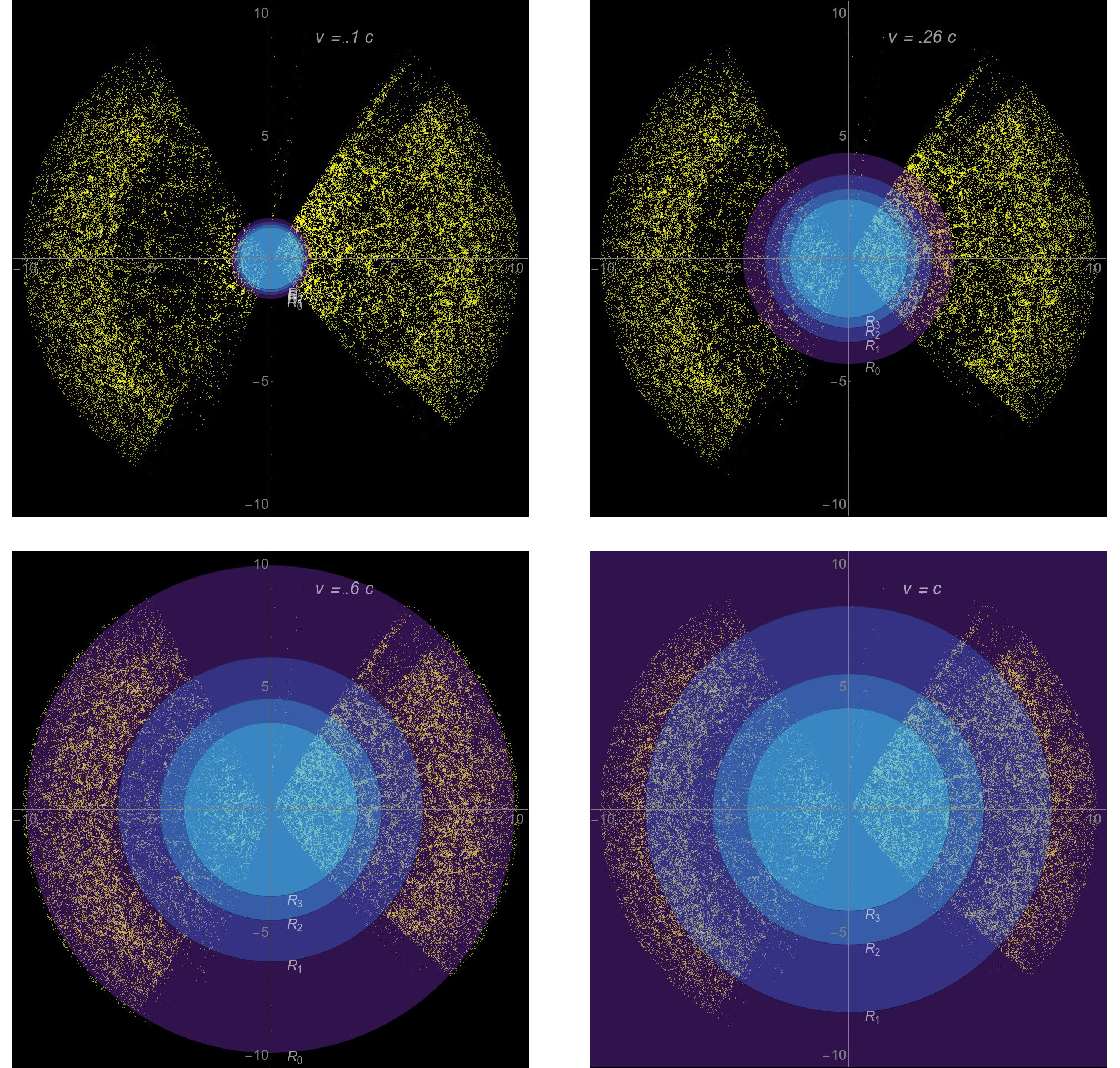}
	\caption{The final domain of an expanding cosmological civilization for four expansion velocities, with boundaries $R_0$, $R_1$, $R_2$, and $R_3$ labeled. Domains are overlayed onto a slice of galaxy position data from the Sloan Digital Sky Survey~\cite{Ahn2014}, for perspective.}
\end{figure}

\end{widetext}

\section{Discussion and conclusions: The value of settlements beyond the horizon}

The prospect of establishing settlements in the $n=0$ region beyond $R_1$, where they cannot signal the home galaxy, immediately raises a question. Why initiate settlement of such a region, if one can never hope to see it occur? Thinking along these lines inevitably leads to more general questions about the value of intergalactic expansion.

To illustrate two different value scenarios that would disagree on the question of expanding into the $n=0$ region, we first consider an argument due to Bostrom~\cite{bostrom2014}. He describes how cosmic expansion could be initiated by a superintelligence, even if it were entirely focused on a local outcome within its home galaxy:

\begin{quotation}
For example, even if a superintelligence's final goals only concerned what happened within some particular small volume of space, such as the space occupied by its original home planet, it would still have instrumental reasons to harvest the resources of the cosmos beyond. It could use those surplus resources to build computers to calculate more optimal ways of using resources within the small spatial region of primary concern. It could also use the extra resources to build ever more robust fortifications to safeguard its sanctum. Since the cost of acquiring additional resources would keep declining, this process of optimizing and increasing safeguards might well continue indefinitely even if it were subject to steeply diminishing returns.
\end{quotation} 

In this kind of scenario, focused entirely on a local outcome, the returns for further cosmic expansion would in fact drop to exactly zero beyond $R_1$. Thus, the majority (up to 7/8 of the final possible domain volume) would be worthless to the concerns of such a superintelligence. Zero benefit and non-zero cost would presumably mean a halt to expansion beyond $R_1$.

The above scenario is a very specific motive for expansion, and such behavior requires tight coherence of value over cosmic time and distance. On the other hand, eternal expansion \emph{would} be the result if each colonized galaxy \emph{independently} found its own value in settling the nearest uninhabited galaxies. Such behavior would also be robust against mutations --- it matters not if values and goals drift over cosmic time, only that \emph{some} value is expected by initiating the next step of expansion.

In closing, we notice a parallel between settlements in the $n=0$ region and the traditional view of distant human descendants one can never hope to meet. The decision to have children is often motivated by the instrumental goals of the parents, e.g. to be supported in old age. But great-great-great grandchildren have no such instrumental value to their great-great-great grandparents. Nevertheless, there is a compelling desire to contribute to a better world, leaving a legacy for one's posterity. In other words, our descendants have intrinsic value to us, even if we can never meet them. So too might an ECC beyond $R_1$. Though we cannot hope to witness it, it may be the most important part of humanity's cosmic legacy. 

\begin{acknowledgements}
	
I am grateful to Toby Ord and Anders Sandberg, for discussions regarding the value of settlement beyond the horizon.
	
\end{acknowledgements}

\bibliography{ref5}{}

\end{document}